\documentclass[aps,showpacs,preprintnumbers,amsmath, amssymb]{revtex4}

\usepackage{float}
\usepackage{graphics,epsfig}
\usepackage{graphicx}
\usepackage{dcolumn}
\usepackage{bm}

\begin{document}
\baselineskip=0.8 cm

\title{{\bf On the thermodynamics of the black hole and hairy black hole transitions in the asymptotically flat spacetime with a box}}
\author{Yan Peng$^{1}$\footnote{yanpengphy@163.com},Bin Wang$^{2,3}$\footnote{wang\underline{~}b@sjtu.edu.cn}, Yunqi Liu$^{4}$\footnote{liuyunqi@hust.edu.cn}}
\affiliation{\\$^{1}$ School of Mathematical Sciences, Qufu Normal University, Qufu, Shandong 273165, China}
\affiliation{\\$^{2}$ Center for Gravitation and Cosmology, College of Physical Science
and Technology, Yangzhou University, Yangzhou 225009, China}
\affiliation{\\$^{3}$ Center of Astronomy and Astrophysics, Department of Physics and Astronomy,
Shanghai Jiao Tong University, Shanghai 200240, China}
\affiliation{\\$^{4}$ School of Physics, Huazhong University of Science and Technology, Wuhan, Hubei 430074, China}

\vspace*{0.2cm}
\begin{abstract}
\baselineskip=0.6 cm
\begin{center}
{\bf Abstract}
\end{center}

We study the asymptotically flat quasi-local black hole/hairy black hole model with nonzero mass of the scalar filed.
We disclose effects of the scalar mass on transitions in a grand canonical ensemble
with condensation behaviors of a parameter $\psi_{2}$, which is similar to approaches in holographic theories.
We find that more negative scalar mass makes the phase transition easier to happen.
We also obtain an analytical relation $\psi_{2}\varpropto(T_{c}-T)^{1/2}$ around the critical phase transition points
implying a second order phase transition.
Besides the parameter $\psi_{2}$, we show that metric solutions can be used to disclose properties of transitions.
In this work, we observe that phase transitions in a box are strikingly similar to holographic
transitions in the AdS gravity and the similarity provides insights into holographic theories.

\end{abstract}

\maketitle
\newpage
\vspace*{0.2cm}

\section{Introduction}

As is well known, the asymptotically flat schwarzschild black holes usually have negative specific
heat and thus can not be in equilibrium with the thermal radiation environment.
In order to overcome this problem, York and other authors provided a simple way of enclosing
the black hole in a box \cite{York,Braden}. It has been found that black holes
could have positive specific heat in this quasi-local ensemble and are thermodynamically stable for certain range of parameters.
On the other side, since the AdS boundary could play a role
of the box boundary condition in a sense, the AdS black hole is usually stable \cite{Hawking}.
According to the AdS/CFT correspondence \cite{Maldacena,S.S.Gubser,E.Witten},
holographic superconductors constructed in the
AdS spacetime have attracted a lot of attentions \cite{S. A. Hartnoll-1}-\cite{Shao-Feng Wu}.
Partly with the interest of holography,
there are literatures paying attention to the
similarity between the asymptotically flat gravity with box boundary and
the AdS gravity.

For anti-de Sitter space, the phase structure of charged black holes has served as
a valuable test of the AdS/CFT correspondence.
Compared with the AdS gravity, similar phase structures and same critical transition exponents have been found in the
asymptotically flat gravity with a box boundary and it was argued that the holography first discovered in the AdS spacetimes may
not be only limited to happen there \cite{S. Carlip,Andrew,Lu}.
The similarity of phase transitions observed in the quasi-local gravity may
cast light on proposals for finite-volume holographic theories.
It is more interesting to further disclose similarities in the AdS gravity and the asymptotically flat gravity with a box boundary.

The Einstein-Maxwell systems with box boundary conditions were studied in \cite{Robert,P. Hut,Gibbons} and
it was shown that overall phase structures of these gravity systems
are similar to those of the AdS gravity.
As a further step, it is interesting to generalize the system to a more completed transition model
by including an additional scalar field.
On the other side, it is also meaningful to study the similarity between transitions of this Einstein-Maxwell-scalar system
in a box and those of the s-wave holographic superconductors constructed with scalar fields
coupled to Maxwell fields in AdS background \cite{Robert,P. Hut,Gibbons}.
Recently, P. Basu, C. Krishnan and P.N.B. Subramanian initiated a
thermodynamical study of such Einstein-Maxwell-scalar systems on asymptotically flat background with reflecting
mirror-like boundary conditions for the scalar field \cite{Pallab Basu}.
For certain range of parameters, this model admits stable hairy black hole solutions,
which provides a way to evade the flat space no-hair theorems.
Another important conclusion is that the overall phase structure
of this gravity system in a box is strikingly similar to
that of holographic superconductor systems in the AdS gravity \cite{Basu,Gary T.Horowitz-2}.
By choosing different values of the scalar charge, scalar mass and St$\ddot{u}$ckelberg mechanism parameters,
we found that effects of model parameters in this flat space/boson star system
are qualitatively the same to those in the holographic insulator/superconductor model \cite{YanP,Gary T.Horowitz-2}.
Moreover, we also showed that operators on the box boundary can be used to detect properties of
the bulk transitions and for the second order transitions, there exists a characteristic exponent
in accordance with cases in AdS gravity systems.
So it is interesting to extend the discussion in the horizonless spacetime in \cite{YanP}
to the background of black holes, which have been studied a lot in holographic theories.
On the other hand, it is also meaningful to generalize the quasi-local black hole/hairy
black hole transition model in \cite{Pallab Basu} by considering nonzero mass of the scalar filed
since the mass usually plays a crucial role in determining properties of transitions.

It was shown in holographic superconductor theories that more negative mass corresponds to a larger
holographic conductor/supercconductor phase transition temperature
or smaller mass makes the transition more easier to happen \cite{Gary T.Horowitz-3,Q. Pan,Yan Peng-1},
so it is interesting to go on to compare effects of the scalar mass
on transition in a box and those of holographic
superconductor transitions constructed in the AdS gravity.
On the other side, it was shown in \cite{Nicolas} that hairy black holes in a box
can be formed dynamically through the superradiant procedure and
effects of scalar mass on dynamical stability of phases have been investigated in \cite{Dolan,Supakchai}.
So it is also interesting to further study thermodynamical properties of such quasi-local black hole/hairy black hole system
with nonzero scalar mass and disclose how the scalar mass could affect the thermodynamical phase transitions.

This paper is organized as follows. In section II, we introduce the black hole/hairy black hole model in a box
with nonzero mass of the scalar field away from the probe limit. In section III,
we manage to use condensation behaviors of a parameter to disclose properties of
phase transitions with various
scalar mass in a grand canonical ensemble. The last section is devoted to conclusions.

\section{Equations of motion and boundary conditions}

We study the formation of scalar hair on the background of four dimensional asymptotically flat
spacetime in a box. In this paper, we choose a fixed radial coordinate $r=r_{b}$ as the time-like box boundary.
And the corresponding Einstein-Maxwell-scalar Lagrange density reads
\cite{Pallab Basu}:
\begin{eqnarray}\label{lagrange-1}
\mathcal{L}=R-F^{MN}F_{MN}-|\nabla \psi-iq A \psi|^{2}-m^{2}\psi^{2},
\end{eqnarray}
where $\psi(r)$ is the scalar field with mass $m^2$ and
$A_{M}$ stands for the ordinary Maxwell field. q is the charge of the scalar field serving as a coupling
parameter between the scalar field and the Maxwell field.
Here, R is the Ricci scalar tensor and $F=dA$.

For simplicity, we would like to consider the scalar field and the Maxwell field
only depending on the radial coordinates as

\begin{eqnarray}\label{symmetryBH}
\psi=\psi(r),~~~~~~~~A=\phi(r)dt.
\end{eqnarray}

Since we are interested in including the matter fields' backreaction
on the background, we assume the ansatz of the geometry of the 4-dimensional
hairy black hole solution in the form
\cite{Pallab Basu}
\begin{eqnarray}\label{AdSBH}
ds^{2}&=&-g(r)h(r)dt^{2}+\frac{dr^{2}}{g(r)}+r^{2}(d\theta^{2}+sin^{2}\theta d\varphi^{2}).
\end{eqnarray}
where the Hawking temperature of the black hole reads
$T=\frac{g'(r_{h})\sqrt{h(r_{h})}}{4\pi}$ and
$r_{h}$ is the horizon of the black hole satisfying $g(r_{h})=0$.

We can obtain equations of motion for matter fields and metric solutions as
\begin{eqnarray}\label{BHpsi}
\frac{1}{2}\psi'(r)^{2}+\frac{g'(r)}{rg(r)}+\frac{q^2\psi(r)^{2}\phi(r)^2}{2g(r)^2h(r)}+\frac{\phi'(r)^2}{g(r)h(r)}-\frac{1}{r^2g(r)}+\frac{1}{r^2}+\frac{m^2}{2g}\psi^2=0,
\end{eqnarray}
\begin{eqnarray}\label{BHpsi}
h'(r)-rh(r)\psi'(r)^2-\frac{q^2r\psi(r)^{2}\phi(r)^2}{g(r)^2}=0,
\end{eqnarray}
\begin{eqnarray}\label{BHphi}
\phi''+\frac{2\phi'(r)}{r}-\frac{h'(r)\phi'(r)}{2h(r)}-\frac{q^2\psi(r)^{2}\phi(r)}{2g(r)}=0,
\end{eqnarray}
\begin{eqnarray}\label{BHg}
\psi''(r)+\frac{g'(r)\psi'(r)}{g(r)}+\frac{h'(r)\psi'(r)}{2h(r)}+\frac{2\psi'(r)}{r}+\frac{q^2\psi(r)\phi(r)^2}{g(r)^2h(r)}-\frac{m^2}{g}\psi=0.
\end{eqnarray}

In order to describe transitions in detail,
we apply the shooting method to integrate these coupled nonlinear differential equations from $r=r_{h}$
to box boundary $r=r_{b}$ to search for the numerical solutions with box boundary conditions.
At the horizon of the black hole $r=r_{h}$, we impose Taylor expansions of solutions as \cite{Pallab Basu}
\begin{eqnarray}\label{InfBH}
&&\psi(r)=aa+bb (r-r_{h})+cc (r-r_{h})^2+\cdots,\nonumber\\
&&\phi(r)=aaa (r-r_{h})+bbb (r-r_{h})^2+\cdots,\nonumber\\
&&g(r)=AA (r-r_{h})+BB (r-r_{h})^2+\cdots,\nonumber\\
&&h(r)=1+AAA (r-r_{h})+\cdots,
\end{eqnarray}
where $aa, bb, \cdots, AAA $ are parameters and the dots denote higher order terms.
Putting these expansions into equations of motion, we could use
three independent parameters $r_{h}$, $aa$ and $aaa$ to describe the solutions.
The scaling symmetry $r\rightarrow \alpha r$ can be used to set $r_{b}=1$. Around the box boundary $(r_{b}=1)$, we assume asymptotic
behaviors of the scalar field and the Maxwell field as
\begin{eqnarray}\label{InfBH}
\psi\rightarrow \psi_{1}+\psi_{2}(1-r)+\cdots,\\
\phi\rightarrow \phi_{1}+\phi_{2}(1-r)+\cdots,
\end{eqnarray}
where $\mu=\phi(1)=\phi_{1}$ is interpreted as the chemical potential. In this paper, we will fix
the chemical potential and work in a grand canonical ensemble.
With the symmetry $h\rightarrow \beta^2 h,~\phi\rightarrow \phi,~t\rightarrow\frac{t}{\beta}$ \cite{Pallab Basu},
we make a transformation to set $g_{tt}(1)=-1$.
Since there are reflecting mirror-like boundary conditions for the scalar field as $\psi(r_{b})=0$,
we fix $\psi_{1}=0$ instead and try to use another parameter $\psi_{2}$ to describe the phase transition,
which is similar to approaches in holographic superconductor theories.
Here, we point out that this box boundary condition $\psi_{1}=0$ is independent
of the scalar mass, which is different from cases in holographic superconductor theories where
asymptotic behaviors of scalar fields at the infinity boundary usually depend on the scalar mass.
And we will show in the following section that $\psi_{2}$
is a good probe to the critical temperature and also the order of transitions in a box.

\section{Properties of phase transitions in a box}

In this part, we firstly plot the numerical solutions as a function
of the radial coordinate with $q=100$, $m^{2}=0$, $\mu=0.15$ and $\psi(r_{h})=0.1$ in Fig. 1.
In this paper, we take $q=100$ as an example for reasons that hairy black holes in a box are
usually only global stable for large charge of the scalar field \cite{Pallab Basu}.
When choosing $m^2=0$, our results are related to the right panel of Fig. 11 in \cite{Pallab Basu}.
It can be easily seen from the left panel of Fig. 1 that this gravity system admits scalar hairy black hole solutions
and at the boundary there is reflecting condition for the scalar field or $\psi(r_{b})=0$.
We also represent behaviors of the metric solutions $h(r)$ in the right panel.
Since we have $h(r)=1$ for cases in the probe limit,
behaviors of curves in the right panel show that the metric is
deformed when considering the matter fields' backreaction on the background.

\begin{figure}[h]
\includegraphics[width=190pt]{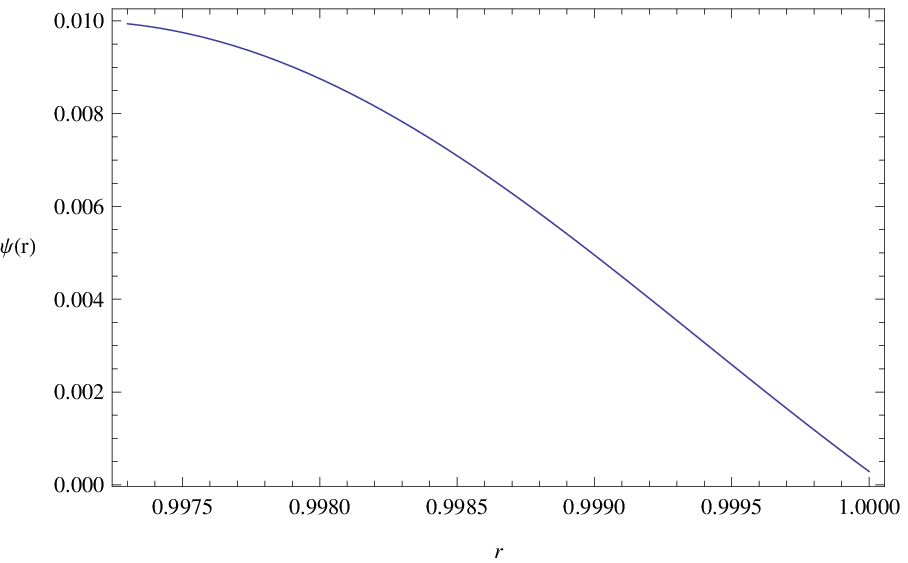}\
\includegraphics[width=190pt]{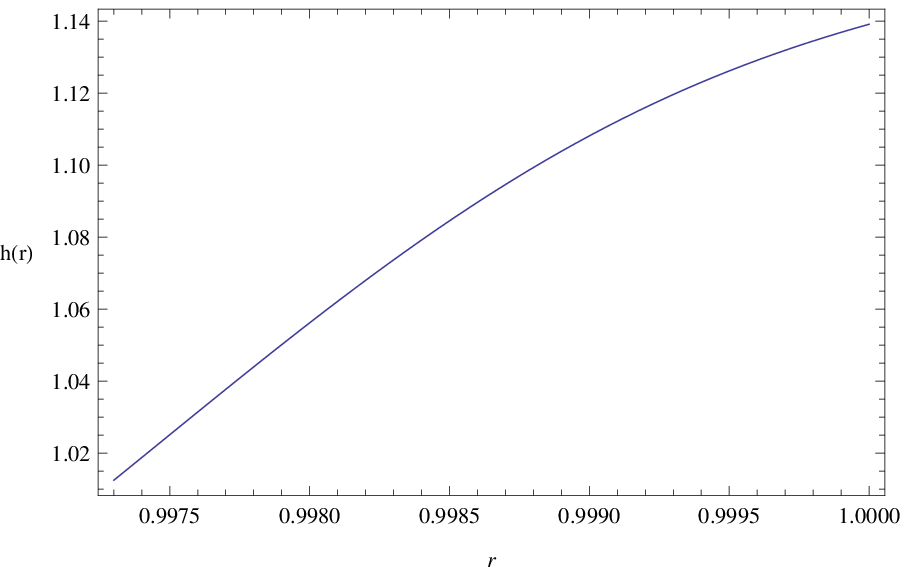}\
\caption{\label{EEntropySoliton} (Color online) We plot solutions as a function of the
radial coordinate $r$ with $q=100$, $m^{2}=0$, $\mu=0.15$ and $\psi(r_{h})=0.1$. The left panel shows behaviors
of the scalar field $\psi(r)$ and the right panel is with values of the metric solution $h(r)$.}
\end{figure}

According to the no-hair theorem, the scalar field cannot attach to the black hole in the usual situation
\cite{Bekenstein,Chase,Ruffini-1,Hod-5,Herdeiro-1,Benone,Brihaye}.
The hairy black holes constructed in AdS spacetimes provide a challenge to the black hole no-hair theorem.
In the AdS gravity, there is an infinite potential wall at the AdS boundary to confine the scalar field \cite{YDB}. The dynamical formation of
scalar hairy black hole due to the AdS boundary was discussed in \cite{PB}.
In the previous discussion, we have obtained scalar hairy black holes in asymptotically flat gravity with a box boundary.
Comparing with the AdS gravity,  the reflecting box boundary  plays the similar role to confine the scalar field, thus
it is natural for the scalar field to condense around a black hole  mimicking the
condensation we observed in the AdS black hole.

We can further see the similarity from the effective potential behavior. Considering a scalar perturbation $\psi$ on the normal charged black hole, $g(r)=1-\frac{2M}{r}+\frac{Q^2}{r^2}$, with charge Q and mass M, the scalar field perturbation can be described in
the Schr$\ddot{o}$dinger-like equation \cite{HF,SH}, where $dr^{*}=\frac{dr}{g(r)}$ and $R(r)=r\psi(r)$,
\begin{eqnarray}
\frac{d^{2}R(r)}{dr^{*2}}=V(r)R(r).
\end{eqnarray}
The effective potential $V(r)=(1-\frac{2M}{r}+\frac{Q^2}{r^2})(m^2+\frac{2M}{r^3}-\frac{2Q^2}{r^4})-\frac{q^2Q^2}{r^2}$, is  a function of the radius coordinate $r$. It can be seen from Fig. 2 that there is no potential well outside the black hole for the scalar field to accumulate.  However, if we
impose a box reflecting boundary (or infinity repulsive potential)
at $r_{b}=1$, we see that a potential well emerges, so that the scalar field can be confined and condensed in the well to lead to the formation of the scalar hair. The formation of hairy black holes with low frequency scalar perturbation if  an additional
box was placed far enough from the black hole horizon was observed in \cite{Nicolas,Dolan}. The appearance of the potential well in the asymptotically flat gravity with a box boundary is similar to that observed in the AdS case and plays the similar role for the scalar filed to condense.

\begin{figure}[h]
\includegraphics[width=200pt]{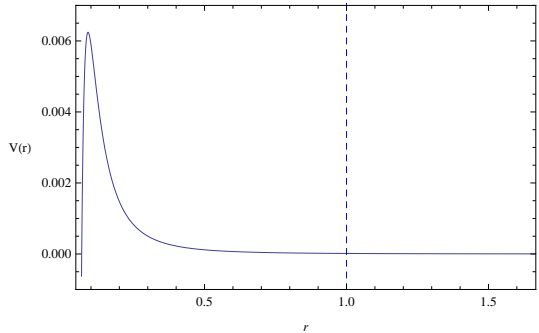}\
\caption{\label{EEntropySoliton} (Color online) We plot $V(r)$ as a function of the
radial coordinate $r$ with $q=100$, $m^{2}=0$, $Q=0.001$ and $M=2$. The vertical
dashed blue line $r_{b}=1$ corresponds to the box boundary and the horizon is around $r_{h}\thickapprox 0.06667$.
Here, we have used the scaling symmetry $r\rightarrow \alpha r$ to set $r_{b}=1$ with $\alpha=\frac{1}{60}$.}
\end{figure}

It is well known that the free energy is powerful in disclosing properties of phase transitions.
The authors in \cite{Pallab Basu} have proposed a way to calculate
the free energy of the system by doing a subtraction of the flat space background.
In this way, the free energy of the Minkowski box is set to be zero.
We show the free energy of the gravity system as a function of the temperature in cases
of $q=100$, $m^{2}=0$ and $\mu=0.15$ in the left panel of Fig. 3.
Since the physical procedure is with the lowest free energy,
we can choose only one phase for every fixed value of the temperature.
It can be easily seen from the left panel that the solid blue line is physical
and there is a critical temperature $T_{c}=0.4910$, below which the normal black hole
phase changes into the hairy black hole phases.
As the free energy is smooth with respect to the temperature around phase transition points,
this black hole/hairy black hole transition is of the second order.

\begin{figure}[h]
\includegraphics[width=190pt]{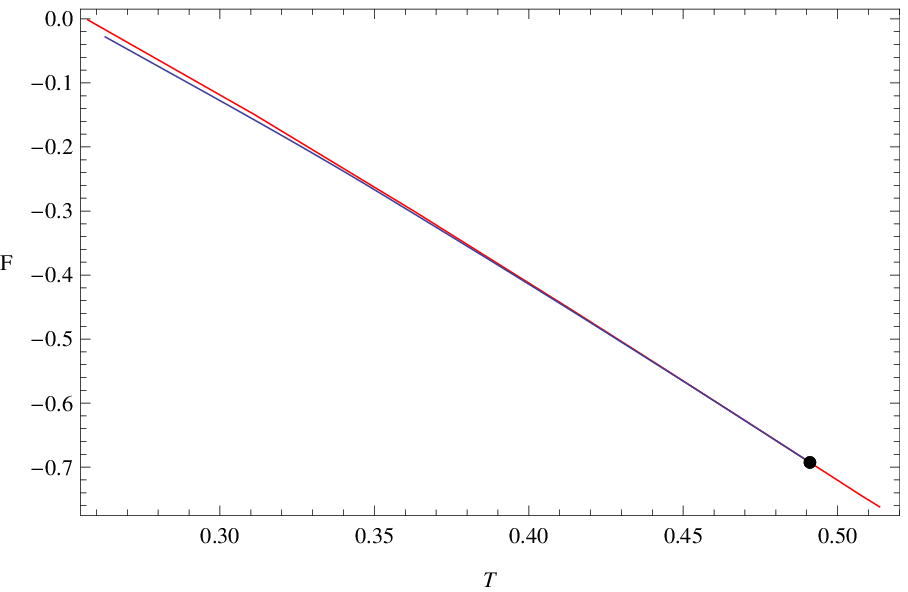}\
\includegraphics[width=190pt]{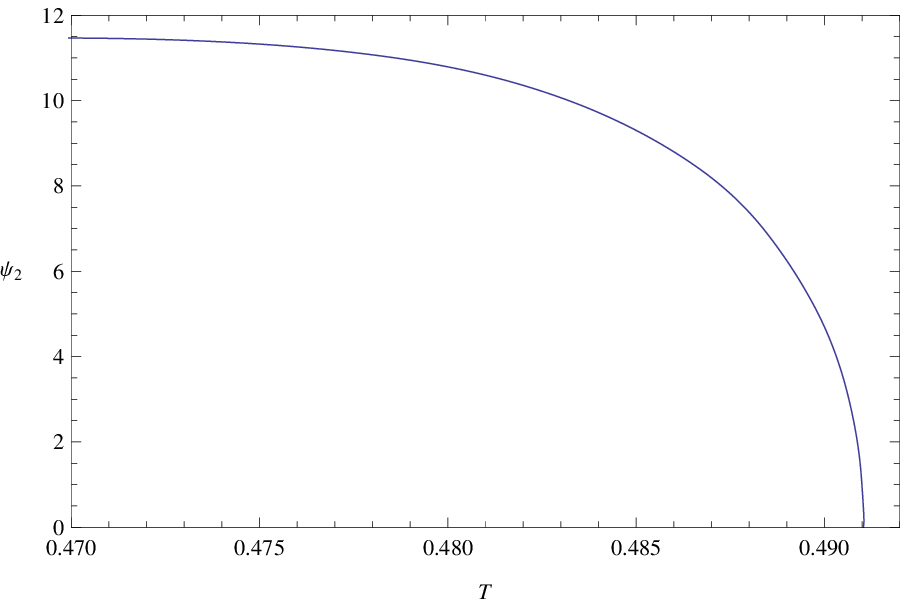}\
\caption{\label{EEntropySoliton} (Color online) We plot the free energy as a function of the temperature
in the left panel with $q=100$, $m^{2}=0$, $\mu=0.15$ and the right panel is with behaviors
of $\psi_{2}$. The solid blue line represents hairy black hole phases and the solid red
line in the left panel shows the free energy of the normal black hole.
We also plot the critical transition phase with a solid black point in the left panel.}
\end{figure}

Inspired by approaches in the holographic superconductor theory, we also want to disclose properties
of transitions from condensation diagram directly related to asymptotical behaviors of the scalar field
on the boundary. In this work, we use $\psi_{2}$ as a parameter to describe properties of phase transitions.
In the right panel of Fig. 3, we plot $\psi_{2}$ with respect to $T$ in cases of $q=100$, $m^{2}=0$ and $\mu=0.15$.
It can be seen from the right panel that $\psi_{2}$ becomes larger as we choose a smaller
temperature, which is qualitatively the same with properties in
holographic conductor/superconductor transitions \cite{T. Nishioka,Gary T.Horowitz-2}.
We also obtain a critical transition temperature $T=0.4910$ in the right panel,
below which the parameter $\psi_{2}$ becomes nonzero.
We mention that this critical transition temperature $T=0.4910$ is
equal to $T_{c}=0.4910$ from behaviors of the free energy in the left panel.
As a summary, we conclude that the parameter $\psi_{2}$ can be used to
determine the critical temperature of the black hole/hairy black hole transition in a box.

\begin{figure}[h]
\includegraphics[width=200pt]{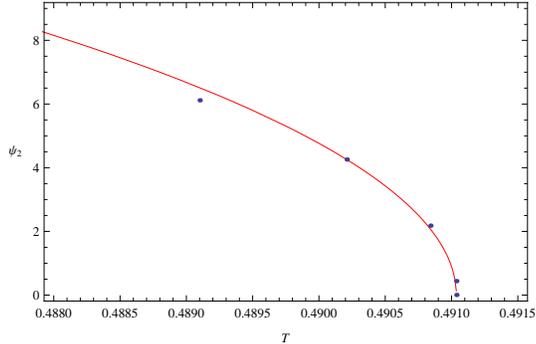}\
\caption{\label{EEntropySoliton} (Color online) We plot hairy black hole phases with solid
blue points. And the solid red line corresponds to the function of $\psi_{2}= 148(T-T_{c})^{1/2}$
with $T_{c}=0.4910$.}
\end{figure}

By fitting the numerical data around the phase transition points,
we arrive at an analytical relation $\psi_{2}\varpropto (T_{c}-T)^{\beta}$ with $\beta=\frac{1}{2}$,
which also holds for the condensed scalar operator
in the holographic conductor/superconductor system in accordance with mean field theories signaling
a second order phase transition \cite{S. A. Hartnoll-1,S. Franco-2,Q. Pan-1,Yan Peng-4}.
We have plotted the fitting formula $\psi_{2}\thickapprox 148 (T_{c}-T)^{1/2}$ with $T_{c}=0.4910$
in Fig. 4 with red solid line.
It can be seen from the picture that the solid blue points representing hairy black hole phases
almost lie on the solid red line around the critical phase transition temperature.
Comparing with the results in Fig. 3, we make a conclusion that this relation between condensation and temperature is a general property
of second order phase transitions for both AdS gravity and asymptotically flat gravity in a box.
Here, the parameter $\psi_{2}$ plays a role strikingly similar to
the condensed scalar operator in holographic conductor/superconductor theories.

We have used operators on the box boundary
to study the bulk gravity and it works well in revealing the critical phase transition
temperature and also the order of transitions.
We also found a critical exponent $\beta=\frac{1}{2}$ as a characteristic of second order transitions,
which usually holds in holographic transitions in asymptotically AdS gravities.
These properties imply that the operator on the hard cut-off box
boundary covers some information of the bulk transition.
Our discussion here provides additional signatures that boundary/bulk correspondence may exist in asymptotically
flat spacetimes.

\begin{figure}[h]
\includegraphics[width=200pt]{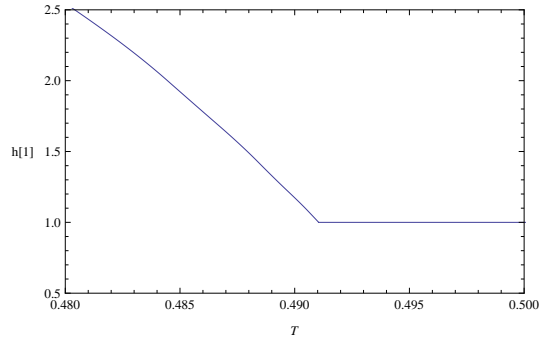}\
\caption{\label{EEntropySoliton} (Color online)
We show behaviors of $h(1)$ as a function of the temperature $T$ with $q=100$, $m^2=0$ and $\mu=0.15$.}
\end{figure}

Besides the condensed scalar operator, it has been found that metric solutions also
can be used to disclose properties of holographic conductor/superconductor transitions
and the jump of metric solutions with respect to the temperature
corresponds to a second order phase transition \cite{Yan Peng-4}.
As a further step, we plan to examine whether the metric solutions can be used to study
transitions in this quasi-local black hole/hairy black hole transition model.
We show behaviors of $h(1)$ as a function of the temperature in Fig. 5 with $q=100$, $m^2=0$ and $\mu=0.15$.
It can be seen from the picture that
$h(1)$ has a jump of the slope with respect to the temperature at the critical phase transition
points $T_{c}=0.4910$ implying second order phase transitions in accordance with results in Fig. 3. We conclude that
metric solutions can be used to disclose the threshold phase transition temperature
and the order of transitions in asymptotically flat black hole/hairy black hole systems in a box,
which is similar to properties of holographic conductor/superconductor transitions in AdS gravity \cite{Yan Peng-4}.

\begin{figure}[h]
\includegraphics[width=190pt]{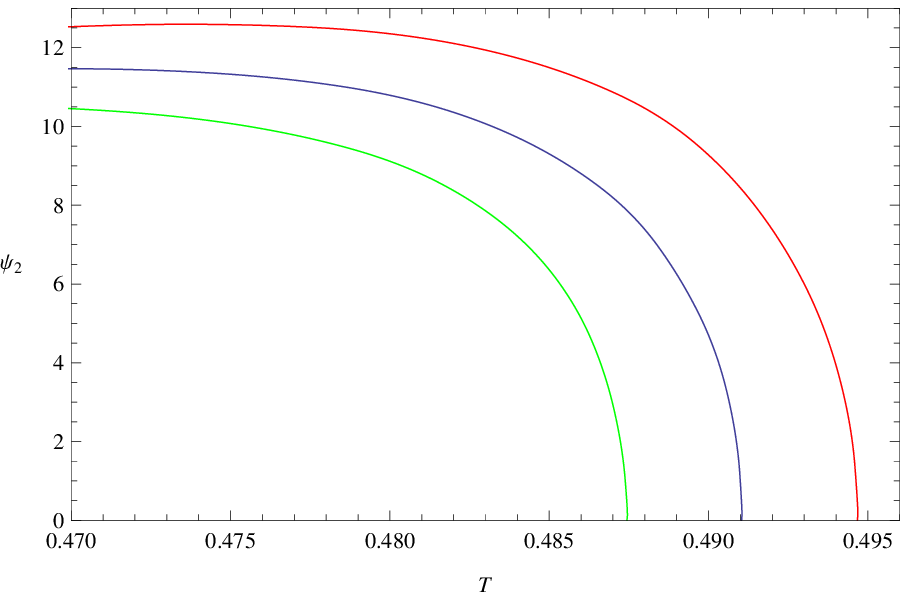}\
\includegraphics[width=190pt]{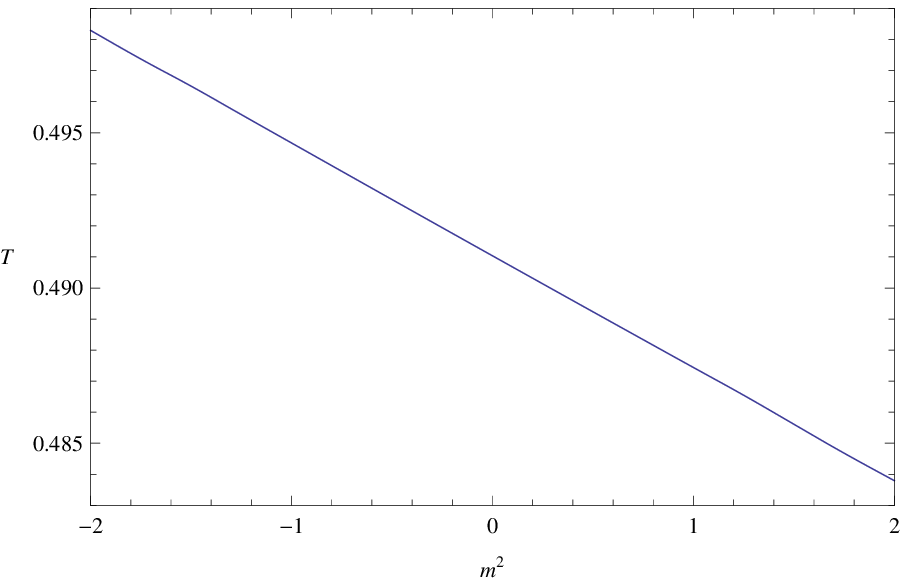}\
\caption{\label{EEntropySoliton} (Color online)
In the left panel, we plot the parameter $\psi_{2}$ with respect to the temperature
with $q=100$, $\mu=0.15$ and various scalar mass $m^2$ from left to right as: $m^2=-1$ (red), $m^2=0$ (blue) and $m^2=1$ (green).
In the right panel,we show the critical temperature $T_{c}$ as a functions of the scalar mass $m^{2}$ with $q=100$ and $\mu=0.15$.}
\end{figure}

For every set of parameters, we obtain a critical temperature $T_{c}$,
below which normal black hole phases transform into hairy black hole phases.
By choosing $q=100$, $\mu=0.15$ and various scalar mass $m^2$,
we disclose how the scalar mass could affect the critical temperature $T_{c}$ in the left panel of Fig. 6.
We can easily see from curves in the left panel that $T_{c}$ decreases as we choose a larger $m^2$.
With more detailed calculations, we go on to plot $T_{c}$
as a function of the scalar mass in the right panel of Fig. 6.
We again arrive at a conclusion that $T_{c}$ becomes smaller as we choose a less negative scalar mass or larger mass
makes the phase transitions more difficult to happen, which is qualitatively similar to cases in holographic transitions in AdS gravity.

\section{Conclusions}

We studied a general four dimensional black hole/hairy black hole transition model in a box
with nonzero mass of the scalar field in a grand canonical ensemble.
Similar to approaches in holographic superconductor theories,
we disclosed properties of transitions through condensation behaviors of a parameter $\psi_{2}$.
With various scalar mass, we examined how the scalar mass can
affect the critical temperature mainly from behaviors of $\psi_{2}$.
We found the more negative scalar mass corresponds to a larger critical temperature and
makes the black hole/hairy black hole transition easier to happen.
In particular, we obtained an analytical relation $\psi_{2}\varpropto(T_{c}-T)^{1/2}$
implying a second order phase transition.
Besides the parameter $\psi_{2}$, we also showed that the metric solutions
can be used to detect the critical phase transition temperature and the order of transitions.
We mention that condensation behaviors of the parameter $\psi_{2}$ is strikingly similar
to those of the scalar operator in holographic transition system and
properties of transitions in this general quasi-local asymptotically flat gravity
are qualitatively the same to those of holographic conductor/superconductor theories in AdS gravity.
In summary, we obtained properties of bulk transitions once also observed in
holography in AdS gravity. Our results provide additional signature that holographic
theories may also exist in quasi-local asymptotically flat gravity.

\begin{acknowledgments}

We would like to thank the anonymous referee for the constructive suggestions to improve the manuscript.
This work was supported by the National Natural Science Foundation of China under Grant Nos. 11305097 and 11505066;
Bin Wang would also like to acknowledge the support by National Basic Research Program of China (973 Program 2013CB834900)
and National Natural Science Foundation of China.

\end{acknowledgments}

\end{document}